\documentclass[conference]{IEEEtran}

\IEEEoverridecommandlockouts                              
\usepackage{graphics}       
\usepackage{epsfig}         
\usepackage{times}          
\usepackage{amsmath}        
\usepackage{amssymb}        
\usepackage{color}

\usepackage{array}

\begin{document}

\title{Impulse-Based Hybrid Motion Control}

\author{\IEEEauthorblockN{Michael Ruderman}
\IEEEauthorblockA{University of Agder (UiA), Post box 422,
4604-Kristiansand, Norway \\ Email: \tt\small
michael.ruderman@uia.no}
\thanks{\textcolor[rgb]{0.00,0.00,1.00}{Preprint of the manuscript accepted to IEEE 43rd Annual
Conference of Industrial Electronics Society (IECON2017)}} }

\maketitle
\thispagestyle{empty}
\pagestyle{empty}

\begin{abstract}
The impulse-based discrete feedback control has been proposed in
previous work for the second-order motion systems with damping
uncertainties. The sate-dependent discrete impulse action takes
place at zero crossing of one of both states, either relative
position or velocity. In this paper, the proposed control method
is extended to a general hybrid motion control form. We are using
the paradigm of hybrid system modeling while explicitly specifying
the state trajectories each time the continuous system state hits
the guards that triggers impulsive control actions. The conditions
for a stable convergence to zero equilibrium are derived in
relation to the control parameters, while requiring only the upper
bound of damping uncertainties to be known. Numerical examples are
shown for an underdamped closed-loop dynamics with oscillating
transients, an upper bounded time-varying positive system damping,
and system with an additional Coulomb friction damping.
\end{abstract}

\section{INTRODUCTION}
\label{sec:1}

Hybrid impulsive control \cite{MillRub03} combines the piecewise
continuous system dynamics with a controlled impulsive behavior,
at which a certain jump in the system state occurs at time
instants of the discrete impulse action. Both the impulsive
control stimulus and, in consequence, resulted state jumps are
triggered by the state trajectory hits some prescribed boundary
region, therefore fulfilling the so-called \emph{guard condition}
within the state-space. Note that in that case the set of guards
constitutes are inherent part of the hybrid control law. Impulsive
control obviously belongs to a general framework of the hybrid
control systems and, more specifically, provides the system
excitation (correspondingly actuation) in form of the infinitely
short impulsive stimuli. Despite the methodologies and theoretical
examples of the hybrid impulsive and, more generally, switching
controls are well-known from the literature e.g.
\cite{Yang01,MillRub03,Guan05} along with fundamentals on the
hybrid dynamics and hybrid control systems
\cite{Branicky98,DeCarlo00,haddad2006impulsive,Goebel09}, their
applications, in particular in motion control, remain rather
modest than one might expect. Not least, a nontrivial analysis and
complexity of the associated design methods are responsible for a
gap between the theoretical framework of hybrid control systems
and their entrance into the engineering practice. In addition, a
whole slew of the practical implementation issues, for details we
refer to an extensive tutorial \cite{Goebel09}, make the hybrid
control systems to a challenging task for the applications.

Few examples of applying the impulsive control strategies to the
motion systems, found in the published works, are mentioned below.
An explicit consideration of impulsive control for the motion
control systems with uncertain friction has been made in
\cite{van2012robust}. The authors considered a set-point
stabilization for the class of position-, velocity-, and
time-dependent friction laws with uncertainty, thus extending an
impulsive control strategy originally proposed in
\cite{orlov2009impulsive}. Both approaches
\cite{orlov2009impulsive} and \cite{van2012robust} apply the
controlled impulsive forces when the system ``gets stuck'' at
non-zero steady-state control error, i.e. when reaching zero
velocity in a bounded vicinity to the set-point. Also the use of
mechanical stroke impulses for precise alignments of components in
the precision engineering and optics has been reported in
\cite{siebenhaar2004precise}, however without an explicit analysis
and design of the control methods. Some neighboring strategies of
controllers with a sequence of adapted/modulated pulses, also
referred to as dithering, have been former used in the
applications, e.g. for positioning tables \cite{yang1987adaptive}
and pneumatic control valves \cite{hagglund2002friction}. Also the
experimental evaluation of impulse-based control, similar as
addressed in this paper, has been previously shown for the linear
guidance drive system in \cite{ruderman2014impulse}.

The aim of this paper is to elaborate the impulsive control method
originally proposed in \cite{ruderman2014impulse} to a general
hybrid motion control formulation suitable for the actuated
systems with time-varying and uncertain damping characteristics.
The outline of the paper is as follows. In Section II, we briefly
summarize the \emph{impulse-based discrete feedback control},
while adapting the control law to be the function of the states
rather than time. The hybrid motion control system, in the sense
of an \emph{autonomous-impulse hybrid system} \cite{Branicky98},
is formulated in Section III while combining the standard
proportional-derivative feedback regulator with the proposed
impulsive control action aimed for reaching the set-point position
despite the damping uncertainties. We provide an explicit analysis
of trajectory solutions at zero crossing impulsive control actions
and derive the corresponding jump map for the hybrid inclusions.
The efficiency of the proposed hybrid motion control is
demonstrated in Section IV by three different numerical examples.
Finally, some concluding remarks are drawn in Section V.

\section{IMPULSE-BASED DISCRETE FEEDBACK CONTROL}
\label{sec:2}

The impulse-based discrete feedback control has been proposed in
\cite{ruderman2014impulse} for the motion systems with damping
uncertainties. The latter assume a relative displacement $x$ with
the bounded uncertain damping coefficient $d$ and the known mass
$m$. The time-varying damping is driven by some unknown
disturbance $\xi(t)$ so that
\begin{equation}
0 < \underline{d} \leqslant d(\xi)\leqslant\overline{d}.
\label{eq:e1}
\end{equation}
The lower and upper bounds of the damping coefficient are
indicated by the subscript and superscript correspondingly. Only
the upper bound is supposed to be known. The impulse-based
discrete feedback control law, similar to the one which is
previously used in \cite{ruderman2014impulse}, is given by
\begin{equation}
u=-\alpha  \, \mathrm{sign}(\dot{x})
\frac{\mathrm{d}}{\mathrm{d}x}\, \mathrm{sign}(x)- \beta \,
\mathrm{sign}(x) \frac{\mathrm{d}}{\mathrm{d}
\dot{x}}\mathrm{sign}(\dot{x}), \label{eq:e2}
\end{equation}
where $\alpha, \, \beta>0$ are the design parameters. Here and
further on we will write the dynamic system quantities without
explicit time argument, this for the sake of simplicity. Note that
unlike in \cite{ruderman2014impulse}, the introduced control law
(\ref{eq:e2}) utilizes the sign derivatives with respect to the
states and not time argument. This allows a direct relation to the
Dirac delta-function used further on in Section \ref{sec:3} for
analysis. Furthermore, we recall that despite the sign operator is
not differentiable at zero, in the ordinary sense, its derivative
under the generalized notion of differentiation in the
distribution theory is twofold of the Dirac
$\delta(\cdot)$-function. The later satisfies the identity
property of integration over the argument
$$\int \delta(y) dy=1,$$
which we will make use of when analyzing later in Section III the
weighted control actions of (\ref{eq:e2}).

The impulsive control action occurs each time the state trajectory
crosses one of the state axes so that both summands in
(\ref{eq:e2}) are disjunctive. Note that they act simultaneously
in zero equilibrium only. Though, when reaching zero equilibrium,
the overall control action becomes apparently zero, that follows
from the limiting conditions $(x,\dot{x})\rightarrow 0$
substituted into (\ref{eq:e2}) for both possible configurations in
the phase-plane, $(x<0, \,\dot{x}>0)$ and $(x>0,\,\dot{x}<0)$. By
implication, for any instantaneous point within the state-space
except zero equilibrium, both impulsive control summands in
(\ref{eq:e2}) can be analyzed independently. In other words, they
discrete impulsive execution is joint by the logical operator OR.
It is also worth noting that the discrete control system switched
at the states zero crossing, also denoted as ``twisting
algorithm'' in \cite{fridman2015}, has been former proposed in
\cite{emelyanov86} and later analyzed on the finite time
convergence and robustness in \cite{orlov2004}.

\section{HYBRID MOTION CONTROL SYSTEM}
\label{sec:3}

\subsection{Feedback controlled motion dynamics}
\label{sec:3:sub:1}

Now, consider a hybrid motion control system
\begin{equation}
m\ddot{x} + d(t)\dot{x} = u + v,
\label{eq:e3}
\end{equation}
with the time-varying damping complying (\ref{eq:e1}), and the
control input which combines the impulse-based discrete feedback
action (\ref{eq:e2}) and a standard linear feedback regulator $v$.
The choice of the latter depends on the eigen-dynamics of system
under consideration, i.e. left-hand-side of (\ref{eq:e3}). For
instance, without damping uncertainties i.e. $d(t)=\mathrm{const}$
the standard proportional-derivative (PD) control
\begin{equation}
v = K(x_r-x) - D \dot{x} \label{eq:e4}
\end{equation}
ensures an asymptotic position convergence to the set-point $x_r$.
Well-known the feedback control gains $K, \,D > 0$ allow for
arbitrary shaping the second-order closed-loop dynamics. Making
the coordinate's shift to zero set-point, i.e. $x_r=0$, the
original control problem transforms into that of an unforced
motion with non-zero initial condition $x(0) \neq 0$. Substituting
(\ref{eq:e2}) and (\ref{eq:e4}) into (\ref{eq:e3}) results in a
hybrid motion control system
\begin{eqnarray}
\nonumber  m\ddot{x} + \bigl(d(t)+D\bigr)\dot{x} + K  x  = -\alpha
\, \mathrm{sign}(\dot{x})
\frac{\mathrm{d}}{\mathrm{d}x}\, \mathrm{sign}(x)- \\
 - \beta \, \mathrm{sign}(x) \frac{\mathrm{d}}{\mathrm{d} \dot{x}}\mathrm{sign}(\dot{x}).
\label{eq:e5}
\end{eqnarray}

\subsection{Autonomous-impulse hybrid system}
\label{sec:3:sub:2}

Generally, an autonomous hybrid (control) system, with a piecewise
continuous state dynamics (set-valued flow mapping)
$\mathbf{F}:\,\mathcal{C} \rightarrow \mathbb{R}^{n}$ and
impulsive behavior (set-valued jump mapping)
$\mathbf{J}:\,\mathcal{D} \rightarrow \mathbb{R}^{n}$, can be
described by means of the hybrid inclusions
\cite{lunze2009handbook} as
\begin{eqnarray}
\label{eq:e6}
\mathbf{\dot{x}} \in \mathbf{F}(\mathbf{x})  & \hbox{ if } \mathbf{x} \in \mathcal{C}, \\
\mathbf{x}^{+} \in \mathbf{J}(\mathbf{x})  & \hbox{ if }
\mathbf{x} \in \mathcal{D}.
\label{eq:e7}
\end{eqnarray}
Note that the flow set $\mathcal{C}$ and jump set $\mathcal{D}$
should be disjoint in the state-space, i.e. $ \mathcal{C} \cap
\mathcal{D} = \oslash$. The next state value $\mathbf{x}^{+}$, the
so-called ``successor'', occurs as a consequence of impulsive
control actions which, in conjunction with the system
eigen-dynamics, determine the jump mapping $\mathbf{J}$.

Obviously, the hybrid motion control system (\ref{eq:e5}), with
the state vector $\mathbf{x}=(x,\dot{x})^T$, incorporates only one
single-valued flow map so that
\begin{equation}
\mathbf{F} = \{ f \} = \left(%
\begin{array}{c}
  \dot{x} \\[2mm]
  -\bigl(d(t)+D\bigr)/m \,\dot{x} - K/m \,x \\
\end{array}%
\right). \label{eq:e8}
\end{equation}
Therefore the flow mapping (\ref{eq:e6}) can be released from the
differential inclusion, and the initial hybrid system model can be
transformed into
\begin{eqnarray}
\label{eq:e9}
\mathbf{\dot{x}} = f(\mathbf{x})  & \hbox{ if } \: \mathbf{x} \in \mathcal{C}, \\
\mathbf{x}^{+} \in \mathbf{J}(\mathbf{x})  & \hbox{ if } \:
\mathbf{x} \in \mathcal{D}.
\label{eq:e10}
\end{eqnarray}
Note that, at the same time, the jump inclusion (\ref{eq:e10})
remains the same as in (\ref{eq:e7}) since the jump mapping
incorporates more than one vector-valued function, this according
to the right-hand side of (\ref{eq:e5}). Recall that the latter
provides an impulsive control stimulus each time the state
trajectory hits or crosses one of the state-space axes. This
implies
$$
\mathcal{D} = X_0 \cup \dot{X}_0 \quad \hbox{with} \quad X_0 := \{
\mathbf{x} \: | \: x=0 \}, \; \dot{X}_0 := \{ \mathbf{x} \: | \:
\dot{x}=0 \},
$$ while consequently $\mathcal{C} = \mathbb{R}^{2} \setminus
\mathcal{D} $.

In order to derive the jump map of the hybrid motion control
system (\ref{eq:e5}), consider the corresponding dynamics for two
disjunctive state configurations at zero crossing, i.e. for
$J\bigl((0,\dot{x})^T\bigr)$ and $J\bigl((x,0)^T\bigr)$.
Obviously, for solving the state trajectories and deriving, based
thereupon, the jump mapping we can use the linear state-space
notation of the system matrix and input coupling vector
\begin{equation}
A = \left(%
\begin{array}{cc}
  0 & 1 \\ [2mm]
  -\frac{K}{m} & -\frac{d+D}{m}
\end{array}%
\right), \quad B = \left(%
\begin{array}{c}
  0 \\ [2mm]
  \frac{1}{m}
\end{array}%
\right), \label{eq:e11}
\end{equation}
given the left-hand-side of (\ref{eq:e5}). The general trajectory
solution of (\ref{eq:e5}) can be then written as
\begin{equation}
\mathbf{x}(t) = \exp(A t)\,\mathbf{x}_0 + \int\limits_{t_0}^{t}
\exp\bigl(A (t-\tau)\bigr) B z(\tau) d\tau. \label{eq:e12}
\end{equation}
Here the right-hand-side of (\ref{eq:e5}) is summarized by $z(t)$
and an initial state $\mathbf{x}_0$ is given for time $t_0$.
Further one should recall that the matrix exponential function is
defined as a power series
$$
\exp(At) = \sum\limits_{v=0}^{\infty} \frac{(At)^v}{v \,!}.
$$

It can be shown that for position zero crossing (further denoted
with superscript ``$0x$'') the general solution (\ref{eq:e12})
transforms into
\begin{eqnarray}
\label{eq:e13}
\mathbf{x}^{0x}(t) & = & \exp(A  t) \left(%
\begin{array}{c}
  0 \\
  \dot{x}_0 \\
\end{array}%
\right) + \\
\nonumber & + & \int\limits_{t_0}^{t} \exp\bigl(A (t-\tau)\bigr)
B\, \mathrm{sign}\bigl(\dot{x}(t)\bigr) (-2\alpha) \delta(\tau)
d\tau.
\end{eqnarray}
Note that the inhomogeneous part of solution (\ref{eq:e13}), i.e.
impulse-excited, is sign-specific depending on the motion
direction prior to position zero crossing. Further we note that
due to equivalence $\delta(x) |_{x=0} = \delta(t) |_{t=t_0} = +
\infty$ and $\delta(x) |_{x\neq0} = \delta(t) |_{t>t_0} = 0$ for
the position zero crossing $x(t_0)=0$ we use the $\delta$-impulse
of time argument $\tau$ instead of that of the state, as otherwise
required by (\ref{eq:e2}).

Following the same line of argumentation, it can be shown that for
velocity zero crossing (further denoted with superscript
``$0\dot{x}$'') the general solution (\ref{eq:e12}) results in
\begin{eqnarray}
\label{eq:e14}
\mathbf{x}^{0\dot{x}}(t) & = & \exp(A t) \left(%
\begin{array}{c}
  x_0 \\
  0 \\
\end{array}%
\right) + \\
\nonumber & + & \int\limits_{t_0}^{t} \exp\bigl(A (t-\tau)\bigr)
B\, \mathrm{sign}\bigl(x(t)\bigr) (-c\beta) \delta(\tau) d\tau.
\end{eqnarray}
Note that despite the system eigen-dynamics, which is determined
by (\ref{eq:e11}), ensures an asymptotic convergence to zero
equilibrium, the damping uncertainties can lead to a full motion
stop at non-zero position. This is in case of the motion gets
sticking or creeping, at which no efficient velocity zero crossing
can be detected and, therefore, an undesirable premature motion
stop should be taken into account in (\ref{eq:e14}). This case,
the corresponding derivative of the sign operator yields the
single Dirac impulse, instead of the twofold as in case of an
effective zero crossing. The introduced weighting variable $c$
captures this distinction by
\begin{equation}
c = \left\{%
\begin{array}{ll}
    2 & \hbox{if  } \: \ddot{x}(t_0)\neq 0,  \\[1mm]
    1 & \hbox{otherwise.} \\
\end{array}%
\right.
\label{eq:e15}
\end{equation}

For deriving the jump map of the hybrid motion control system,
i.e. determining the system state $\mathbf{x}^{+} =
\mathbf{x}(t^{+})$ (the ``successor'') after jump, take the limit
condition $t \rightarrow t_0$ for which (\ref{eq:e13}) and
(\ref{eq:e14}) can be directly evaluated. Assuming the state
trajectory hits the jump set $\mathcal{D}$ at time $t_0$, for
which the instantaneous state $\mathbf{x}_0$ is known, results in
\begin{equation}
\mathbf{x}^{+} = \left\{%
\begin{array}{ll}
    \mathbf{x}^{0x}(t_0) & \hbox{if  } \: \mathbf{x}(t_0) \in  X_0,  \\[1mm]
    \mathbf{x}^{0\dot{x}}(t_0) & \hbox{if  } \: \mathbf{x}(t_0) \in  \dot{X}_0.
\end{array}%
\right. \label{eq:e16}
\end{equation}
This yields, after evaluating (\ref{eq:e13}) and (\ref{eq:e14}),
the jump mapping
\begin{equation}
\mathbf{J}(\mathbf{x}) = \left\{%
\begin{array}{ll}
    \mathbf{x} - 2 B \, \alpha \, \mathrm{sign}(\dot{x})   & \hbox{for } \: \mathbf{x} \in  X_0,  \\[1mm]
    \mathbf{x} - c B \, \beta \, \mathrm{sign}(x) & \hbox{for } \: \mathbf{x} \in  \dot{X}_0.
\end{array}%
\right. \label{eq:e17}
\end{equation}

\subsection{Impulsive control gains}
\label{sec:3:sub:3}

The selection of impulsive control gains, $\alpha$ and $\beta$,
follows straight to the jump map since the latter determines the
motion system state immediately after an impulsive control
execution.

Consider first the discrete control action at the position zero
crossing, for which the instantaneous state is $\mathbf{x}(t_0) =
[0,\dot{x}_0]$. Obviously, the desired ``successor'' should lie in
the origin, therefore representing attainment of zero equilibrium.
Substituting $\textbf{x}^{+}=\mathbf{0}$ into (\ref{eq:e17}), for
the first case $\mathbf{x} \in  X_0$, and solving with respect to
the control gain $\alpha$, results in
\begin{equation}
 \alpha = 0.5 m |\dot{x}_0|.
\label{eq:e18}
\end{equation}
Note that the above situation represents an ideal case of
instantaneous velocity change, i.e. assuming infinite
accelerations of the moving mass. Since a real system acceleration
is bounded and with an actuation force dynamics, (\ref{eq:e18})
represents rather the lower gain boundary, below which the state
trajectory cannot, even theoretically, reach zero equilibrium
after an impulsive control execution. For determining the upper
gain boundary one can show that in case of
$$
\alpha = m |\dot{x}_0|,
$$
the jump map yields the state ``successor'' $\textbf{x}^{+} = -
\textbf{x}(t_0)$, here theoretically as well i.e. without taking
into account the real system accelerations. That is the
``successor'' state will remain in zero position but accept the
opposite-sign velocity of the same magnitude as before the last
impulsive control action. Consequentially, the resulted (ideal)
trajectory will end up in an infinite time switching cycle between
$\pm \dot{x}_0$ around zero position. The above consideration
allows us to formulate the overall gain criterion as
\begin{equation}
0.5 m |\dot{x}_0| \leq \alpha  <  m |\dot{x}_0|.
\label{eq:e19}
\end{equation}
Note that the same parameter criterion has been suggested in
\cite{ruderman2014impulse} for the case of the unity mass and
without explicit analysis of the hybrid system dynamics and jump
mapping.

Now consider the problem of determining $\beta$-gain for an
impulsive control action at the velocity zero crossing.
Substituting $\textbf{x}^{+}=\mathbf{0}$ into (\ref{eq:e17}), for
the second case $\mathbf{x} \in \dot{X}_0$, one can see that no
direct solution for attaining zero equilibrium exists. This is
since the impulsive, equally as any other, control action is
unmatched with the position state jump, correspondingly dynamics.
Therefore, the single alternative is in providing the jump to a
predefined velocity $\dot{x}^{+}$ at $x(t_0) \neq 0$ which will
afterwards allow reaching zero equilibrium for the trajectory to
be driven by the eigen-dynamics of (\ref{eq:e11}). For the
critically damped eigen-dynamics, i.e. $(d+D)^2=4mK$, the
homogenous solution is given by
\begin{equation}
x(t) = C_1 \exp(\lambda t) + C_2 \exp(\lambda t) t,
\label{eq:e20}
\end{equation}
where the real double-pole is
$$
\lambda = - (d+D)/(2m).
$$
Note that here we write some (nominal) constant damping $d$ which
complies with (\ref{eq:e1}). Further we explicitly underline that
the feedback controlled system should be critically damped by the
assigned derivative feedback control gain $D$. In case the
controlled motion system becomes underdamped, and that through the
time-varying damping behavior $d(t)$, the trajectory will
inevitably hits $X_0$ and an impulsive action of the position zero
crossing employs as described above. On the contrary, if the
controlled motion system becomes overdamped, which implies an
undesirable premature motion stop, then this case should be
captured by the initial requirement on the upper bound of damping
coefficient to be known, cf. with Section \ref{sec:2}. Taking the
time derivative of the homogenous solution (\ref{eq:e20}), which
is
\begin{equation}
\dot{x}(t) = C_1 \lambda \exp(\lambda t) + C_2 \bigl( \lambda t
\exp(\lambda t) +  \exp(\lambda t)  \bigr), \label{eq:e21}
\end{equation}
and solving for the initial values $\mathbf{x}(0)=[x_0, 0]$ one
obtains
\begin{equation}
C_1 = x_0, \quad C_2 = - x_0 \lambda. \label{eq:e22}
\end{equation}
It is obvious that the above constants of the homogenous solution
(\ref{eq:e20}), (\ref{eq:e21}) reflect both the initial position
and system damping and, therefore, bear the signature of the
required initial velocity to be excited through an impulsive
control action. Moreover it should be underlined that both
constants (\ref{eq:e22}) have been determined for the initial
values of the nominal system, i.e. with a known and constant
damping $d$. Therefore, the nominal system will asymptotically
reach zero equilibrium, after velocity zero crossing, even if
neither impulsive control action takes place. Consequentially,
when substituting the constants (\ref{eq:e22}) into
(\ref{eq:e21}), first a zero initial velocity
\begin{equation}
\dot{x}(0) = C_1 \lambda + C_2 = x_0 \lambda - x_0 \lambda,
\label{eq:e23}
\end{equation}
is obtained as expected. However, for an uncertain,
correspondingly time-varying, system damping the first
$\lambda$-term in (\ref{eq:e23}) becomes uncertain,
correspondingly time-varying, while the second one remains fixed
by the $C_2$ initial value. Therefore, the initial velocity
required for ensuring the $\lambda$-uncertain solution
(\ref{eq:e20}) can reach zero equilibrium is not longer zero. As
assumed in Section \ref{sec:2}, it is sufficient to know the upper
bound of the uncertain system damping so that the first
$\lambda$-term in (\ref{eq:e23}) can be computed for the maximal
damping value $\bar{d}$. After that (\ref{eq:e23}) can be
transformed into
\begin{equation}
\dot{x}(0) = - x_0 \frac{\bar{d}+D}{2m} + x_0 \frac{d+D}{2m} = -
x_0 \frac{\bar{d}-d}{2m}.\label{eq:e24}
\end{equation}
It is evident that the required velocity jump depends on the
position at velocity zero crossing, on the one hand, and
difference between the upper bound of the damping coefficient and
its nominal value $d$, on the other hand. Since the latter is
a-priory unknown and, in worst case, can be infinitesimally low at
instant of the velocity zero crossing, the suggested velocity jump
is
\begin{equation}
\dot{x}^{+} = - x_0 \frac{\bar{d}}{2m}. \label{eq:e25}
\end{equation}
We stress that (\ref{eq:e25}) captures the case of maximal system
damping, while for all lower damping values the resulted motion
trajectory will be guaranteed hitting $X_0$. Now, substituting
(\ref{eq:e25}) into (\ref{eq:e17}), for the second case
$\mathbf{x} \in \dot{X}_0$, and solving with respect to the
$\beta$-gain results in
\begin{equation}
\beta = |x_0| \frac{\bar{d}}{2c}. \label{eq:e26}
\end{equation}
It should be noted that in order to capture the case when the
instantaneous acceleration at velocity zero crossing is
unavailable, cf. case difference in (\ref{eq:e15}), the weighting
factor $c=1$ can be continuously used. Recall that this ensures
the $\beta$-gain to be sufficient also in case of the system
sticking, i.e. when a relative motion stops outside of zero
equilibrium.

\section{NUMERICAL EXAMPLES}
\label{sec:4}

The proposed hybrid motion control is demonstrated for two
numerical examples provided below. The simulation setup is
realized using the Simulink$^{\circledR}$ software from MathWorks
Inc, with the set fixed-step (0.0001) $ode3$ solver, while the
trapezoidal method for integration calculus is used.
\begin{table}[!h]
  \renewcommand{\arraystretch}{1.6}
  \caption{Parameters of numerical simulation setup}
  \small
  \label{tab:1}
  \begin{center}
  \begin{tabular} {|p{0.8cm}|p{0.4cm}|p{0.4cm}|p{0.4cm}|p{0.4cm}|p{0.4cm}|p{0.4cm}|p{0.4cm}|p{1.2cm}|}
  \hline \hline
  Param.      &  $m$  &  $K$  &  $D$ &  $F_c$ & $\underline{d}$ &  $\overline{d}$ & $c$ & $\alpha/m/|\dot{x}_0|$ \\
  \hline
  Value          & 0.1 & 10  & 0.5 &  1  & 0.15  & 1.5  &  2  &  0.6 \\
  \hline \hline
  \end{tabular}
  \end{center}
  \normalsize
\end{table}
The control parameters are set according to the developments
provided in Section \ref{sec:3} and all the system simulation
parameters are listed in Table \ref{tab:1}. Note that the
parameters are assumed as normalized (unitless), so that the
computed system states are correspondingly unitless as well.

\begin{figure}[!h]
\centering
\includegraphics[width=0.8\columnwidth]{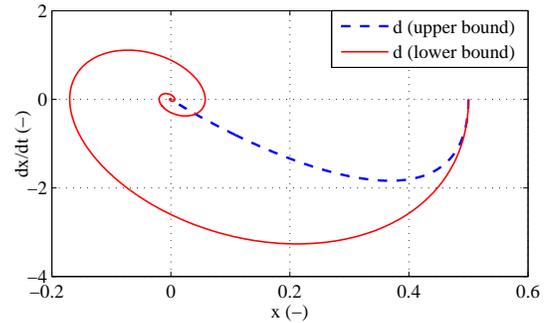}
\caption{Trajectories of the feedback motion control system (5)
without impulsive control part (zero right-hand-side) for initial
value $x(0)=0.5$. The system damping at upper $d=\bar{d}=1.5$ and
lower $d=\underline{d}=0.15$ bounds.} \label{fig:1}
\end{figure}

In order to provide and appropriate sensation of the resulted
system dynamics, the hybrid motion control system (\ref{eq:e5}) is
first simulated without impulsive control action, i.e. with zero
right-hand-side. The non-zero initial position is set to
$x(0)=0.5$ so that the PD feedback control loop forces the state
trajectory to converge towards zero equilibrium. A comparison is
made for two cases -- the lower and upper bounds of the system
damping $d$. The resulted trajectories are shown in Fig.
\ref{fig:1}. Obviously, the case of the upper bound of damping
coefficient represents a critically damped response, cf. with
Section \ref{sec:3:sub:3}. The case of the lower bound of damping
coefficient offers an oscillating trajectory which converges to
zero equilibrium after several periods. Note that the assumed
lower and upper bounds of the system damping differ by an order of
magnitude, see Table \ref{tab:1}.

The above results disclose the reduced performance of the motion
control system in case of the underdamped closed-loop dynamics. In
the following, we are to evaluate the hybrid motion control system
with impulsive control action as in (\ref{eq:e2}).

\subsection{Underdamped closed-loop dynamics}
\label{sec:4:sub:1}

When allowing for the impulse-based control law, i.e. employing
the right-hand side of (\ref{eq:e5}), the hybrid control
performance becomes clearly superior comparing to that
demonstrated above for the PD-controlled motion dynamics without
impulsive control part.
\begin{figure}[!h]
\centering
\includegraphics[width=0.98\columnwidth]{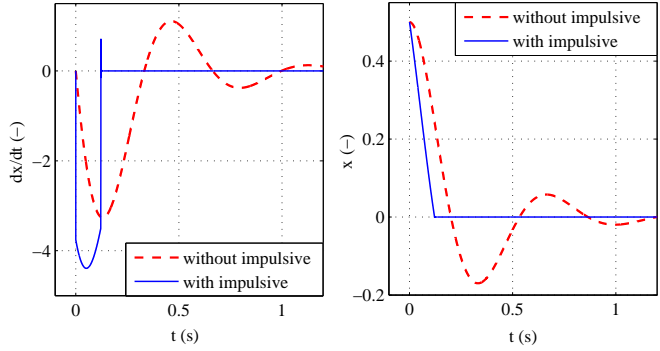}
\caption{Position and velocity response of the feedback motion
control system (5), once with and once without right-hand-side
which is impulsive control action. The plant damping of the low
bound $d=\underline{d}=0.15$ is assumed.} \label{fig:2}
\end{figure}
Here the case of the low bound system damping
$d=\underline{d}=0.15$ is assumed, since that one implies several
periods of the sate oscillations and, consequently, zero crossing
of the state axes at which the impulsive control actions occur.
The relative position and velocity response of both motion control
systems are shown opposite to each other in Fig. \ref{fig:2}.
Obviously, the hybrid motion control provides a much faster
convergence to zero set-position, even with a nearly linear rate
corresponding to the induced velocity. At the same time, the
induced peak velocity is not significantly higher comparing to
that of the pure PD feedback control. Some minor transient peaks
occur in vicinity to zero settling point, but these are fairly
negligible comparing to the transient oscillations of the feedback
control system without impulsive control part.

\subsection{Time-varying damping coefficient}
\label{sec:4:sub:2}

\begin{figure}[!h]
\centering
\includegraphics[width=0.98\columnwidth]{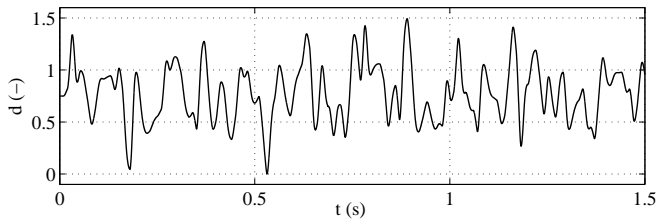}
\caption{Time-varying damping coefficient used for evaluation.}
\label{fig:3}
\end{figure}

A time-varying system damping $d(t)$ is included into the control
system (\ref{eq:e5}) while solely fulfilling the boundary
condition (\ref{eq:e1}). The time-varying damping signal is
generated by using a low-path filtered white-noise with an
additional bias to guarantee that (\ref{eq:e1}) holds. The
simulated time-series of the damping coefficient is shown in Fig.
\ref{fig:3}.
Note that, generally, a time-varying system damping is usual for
e.g. system with mechanical friction, where the weakly-known
internal and external factors render the frictional coefficients
as time- or state-varying and often without an explicit functional
relationship which could assist the control design. For more
details on the kinetic friction uncertainties and their impact on
the linear feedback control systems we refer to \cite{Ruderman15}
and \cite{ruderman2016integral}.

\begin{figure}[!h]
\centering
\includegraphics[width=0.98\columnwidth]{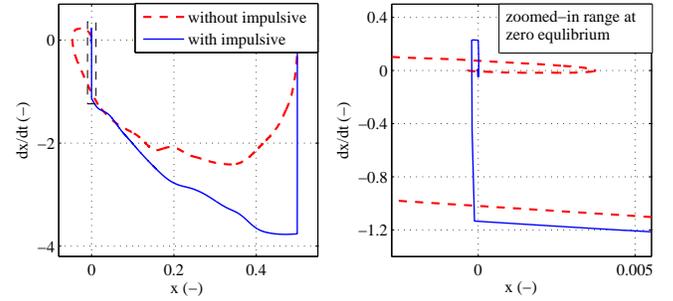}
\caption{Trajectories of the system (5) with time-varying damping,
without and with impulsive control part. The initial value is set
to $x(0)=0.5$.} \label{fig:4}
\end{figure}

The trajectories of the controlled system, once without and once
with the impulsive control part, are compared to each other in
Fig. \ref{fig:4}.
From the zoom-in on the right, it becomes evident that while the
first position zero crossing occurs at the comparable velocities,
a fast (almost immediate) convergence to zero is only in case of
the hybrid impulsive control.

\subsection{System with Coulomb friction}
\label{sec:4:sub:3}

\begin{figure}[!h]
\centering
\includegraphics[width=0.98\columnwidth]{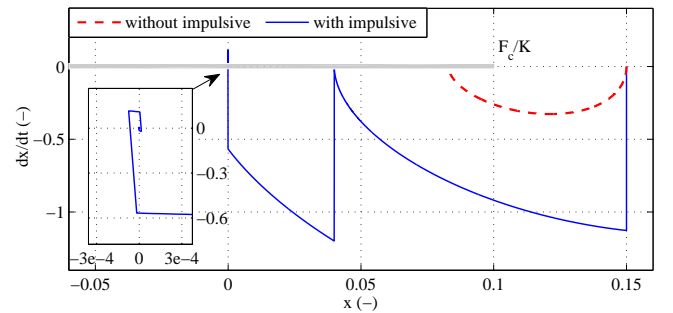}
\caption{Trajectories of the system (5), extended by Coulomb
friction, without and with impulsive control part. The initial
position is set to $x(0)=0.15$.} \label{fig:5}
\end{figure}

The system under control, with the assumed low bound damping
$d=\underline{d}=0.15$, is extended by the Coulomb friction so
that the left-hand-side of (5) results in $m\ddot{x} +
\bigl(d(t)+D\bigr)\dot{x} + K  x + F_c \mathrm{sign}(\dot{x})$.
Therefore, the constant friction force of a magnitude equal to the
Coulomb friction coefficient $F_c > 0$ acts at the unidirectional
motion, thus providing a rate-independent damping until full
motion stop. Well-known, within a certain dead-zone around zero
equilibrium, which is $\pm F_c/K$ for the PD feedback control (4),
the vector field on both sides of the $x$-axis is orthogonal and
points towards $\dot{x}=0$. That is once attaining the dead-zone,
the trajectory stays for always so that neither continuation of
the controlled motion towards zero equilibrium occurs, cf. with
\cite{putra2007}. Accordingly, the interval $-F_c/K \leq x \leq
F_c/K$, see grey bar in Fig. \ref{fig:5}, constitutes the largest
invariant subset of the $x$-axis with an infinite number of stable
equilibria. It implies that convergence of the feedback control
system to zero equilibrium cannot be guaranteed without impulsive
control part, independently of the $K$ and $D$ parameters
selection. The corresponding trajectory is shown in Fig.
\ref{fig:5} for the initial state $[x(0), \dot{x}(0)]=[0.15, 0]$
while the assigned Coulomb friction coefficient results in
$F_c/K=0.1$.
On the contrary, enabling for the impulsive control coaction
allows for reaching zero equilibrium, while an asymptotic
convergence can be guaranteed. It follows due to zero equilibrium
appears as a global attractor as long as $\dot{x}, K \neq 0$.
Recall that the largest invariant subset lies on the $x$-axis, so
that the motion can become `stuck' only at zero velocity. At the
same time, an impulsive control action at hitting the $x$-axis
ensures the velocity jump so that the system temporary releases
from sticking. This is valid even for an infinitesimally short
instant as the position approach zero equilibrium, cf. with (26).
The trajectory which converges to zero equilibrium when enabling
for the impulsive control is equally shown in Fig. \ref{fig:5}.
One can see that after the first impulse, starting from the same
initial position $x(0)=0.15$, the trajectory next reaches the
largest invariant subset. Here the system should experience full
the motion stop when operating without impulsive control coaction.
After hitting the guard, however, the trajectory is repulsed again
and runs towards velocity zero crossing. Afterwards the trajectory
converges to zero equilibrium while being excited each time when
crossing the state axes, see zoom-in in Fig. \ref{fig:5}.

\section{CONCLUSIONS}
\label{sec:5}

In this paper we have introduced the generalized formulation of
the impulse-based hybrid motion control for the second-order
systems with damping uncertainties. Using the unified framework of
the hybrid systems and, more specifically, autonomous-impulse
hybrid systems, we have analyzed the controlled motion dynamics
with the right-hand-side impulsive control actions at both states
zero crossing. An appropriate jump inclusion has been derived for
the state trajectories hit the guards, which are position and
velocity state axes. We have analyzed the ``successor'' state
dynamics, i.e. immediately after execution of impulsive control
actions and, based thereupon, provided the appropriate conditions
for selection of the control gains. Three numerical examples of
the system with (i) significantly underdamped closed-loop
dynamics, (ii) time-varying damping, and (iii) additional Coulomb
friction damping have been demonstrated to emphasize the
performance of the proposed impulse-based hybrid motion control.

\section*{Acknowledgment}

This work has received funding from the European Union Horizon
2020 research and innovation programme H2020-MSCA-RISE-2016 under
the grant agreement No 734832. The author is also grateful to
Prof. Yury Orlov for a helpful discussion on improving the
mathematical rigor of analysis.

\bibliographystyle{IEEEtran}
\bibliography{references}

\end{document}